# A DOZEN NEW DWARF GALAXY CANDIDATES IN THE LOCAL VOLUME

I.D.Karachentsev1), P.Riepe2), T.Zilch2)

1) Special Astrophysical Observatory, Russian Academy of Sciences

2) Tief Belichtete Galaxien group of Vereinigung der Sternfreunde e.V., Fachgruppe Asrtofotografie, PO Box 1169, D-64629, Heppenheim, Germany

**Abstract**

We carried out a survey of low-surface-brightness companions around nine luminous Local Volume galaxies using long exposures with small amateur telescopes. We found 12 low- and very low-surface-brightness objects around the galaxies: NGC 628, NGC 2337, NGC 3368, NGC 3521, NGC 4236, NGC 4258, NGC 4594, NGC 5055, and NGC 6744 situated within 12 Mpc from us. Assuming the dwarf candidates are satellites of the neighbouring massive galaxies, their absolute B magnitudes are in the range of [-8.9, -13.0], linear diameters are [0.6 - 2.7] kpc, and the mean surface brightnesses are [25.1 - 27.5] mag/sq.arecsec. The mean linear projected separation of satellite candidates from the host galaxies is 83 kpc.

*1. Introduction.* The dwarf galaxies, whose luminosity is hundreds of times smaller than the luminosity of the Milky Way, are the dominant population of the Universe. More than half of them are grouped into suites around massive Milky Way type galaxies. Measurements of distances and radial velocities of dwarf satellites relative to their bright central galaxy give a unique opportunity to estimate the total mass of dark matter of the dominant galaxy on scales of ~(200-300) kpc.

Systematic searches for dwarf galaxies over the whole sky [1, 2] based on the POSS-II/ESO-SERC photographic survey led to the discovery of several hundred nearby dwarf systems of predominantly low surface brightness. Follow-up surveys of large areas of the sky in the optical range, SDSS [3], and in the HI-line of neutral hydrogen, HIPASS, ALFALFA [4-7] essentially increased the population of nearby dwarf galaxies. To date, the number of dwarfs in the Local Volume (=LV), restricted by the radius of 11 Mpc, is approaching a thousand. A variety of data on them are collected in the Updated Nearby Galaxy Catalog (=UNGC) [8] and in a regularly updated database [9], http://www.sao.ru/lv/lvgdb.

In recent years, telescopes of the (4-6)-meter class equipped with wide-field CCD cameras and situated in sites with a sub-arcsec seeing have been used to search for ultra-faint dwarf galaxies. Such programs culminated in the discovery of many new satellites around the nearby massive galaxies: M31, M81, M106, M101, and Centaurus A [10-13]. In the case of the nearest galaxies the physical membership of their companions is successful to confirm with resolution of dwarf systems into stars forming the red giant branch.

In parallel with these projects, the search for dwarf galaxies of very low surface brightness with small telescopes is quite successful, too. With long exposures of about 10-50 hours at a ~30-cm telescope, it was possible to detect objects having the mean surface brightness SB ~(26-28) mag/square arcsec and angular dimensions more than 0.2 arcmin [14-17]. A significant role in these efforts belongs to amateur astronomers. One of such successful amateur teams is the TBG group, http://tbg.vdsastro.

*2. TBG team survey.* The TBG (Tief Belichtete Galaxien) group in the Astrophotography department of the German association of VdS was organized by P. Riepe in 2012. The group includes about forty amateurs of astrophotography, who use 10 - 110 cm aperture telescopes equipped with CCD detectors and packages for data processing. One of their main aims is to do long-exposure imaging of nearby bright galaxies. Such deep images are suitable for detecting dwarf satellites around these galaxies (as well as faint stellar streams) with a characteristic surface brightness about (1-2) per cent of the moonless night sky. In recent years, the TBG team discovered almost 30 new dwarf-galaxy candidates in the Local Volume with absolute magnitudes and surface brightnesses typical of the known satellites of Andromeda (M31) and M81. Some of them have been then confirmed as physical companions of bright galaxies via measuring their radial velocities with the 6-meter BTA telescope of SAO RAS. The results of these searches are presented in [16, 18-19]. Below we report on a dozen new dwarf-galaxy candidates discovered around nine bright galaxies in the Local Volume.

*3. Results of searching for dwarf satellites.*

*3.1. NGC 628 = M 74.* This spiral galaxy seen face-on at a distance of 10.2 Mpc [8] has only four known satellites of moderate luminosity: UGC 1056, UGC 1104, UGC 1171, and KDG 10. A new dwarf galaxy with very low surface brightness, NGC628dwTBG, found by us resides to North-West from NGC 628 at the angular separation $r_p$ = 59' that corresponds to the linear projected separation $R_p$ = 174 kpc. Figure 1 reproduces a fragment of the image of the galaxy NGC 628 and its probable new companion that has been derived by M. Blauensteiner.

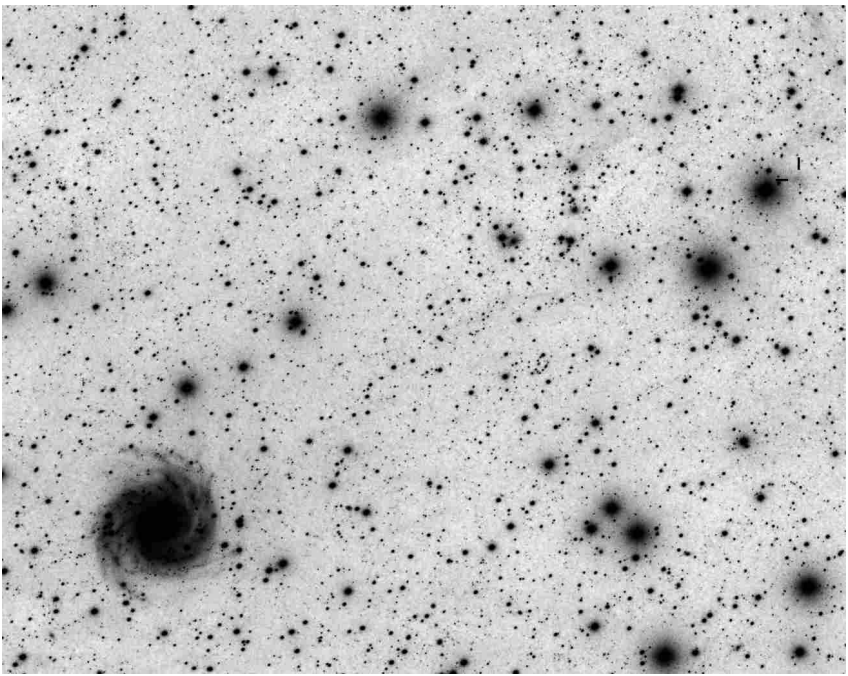

*Fig.1. Spiral galaxy NGC 628 and its supposed new satellite NGC628dwTBG situated in the right-hand top corner. North is up and East is on the left.*

*3.2. NGC 2337 and UGC 3698.* These dwarf galaxies constitute a wide physical pair at a distance of 11.5 Mpc [8] with a radial velocity difference of only 12 km/s. The image of this pair (Figure 2) made by J. Muller reveals a dwarf low-surface-brightness galaxy NGC2337dwTBG1 which probably forms a triple system with pair components.

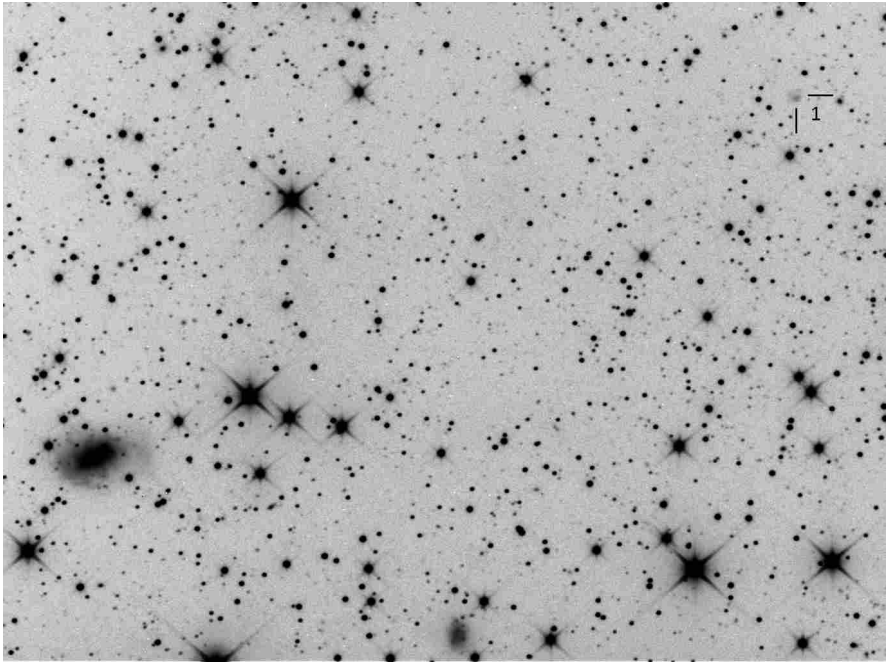

*Fig.2. Galaxy NGC 2337 (at the left-hand side), UGC 3698 (at the bottom), and their new probable companion NGC2337dwTBG1.*

*3.3. NGC 3368= M96.* This Sb-type spiral galaxy at a distance of 10.4 Mpc [8] belongs to the brightest members of the rich group Leo-I, whose dwarf population has been investigated by many authors [20-23]. By now, in the Leo-I group there are more than 30 known dwarf members with low surface brightness. The image of M96 derived by O. Schneider (Figure 3) adds one more new object, NGC3368

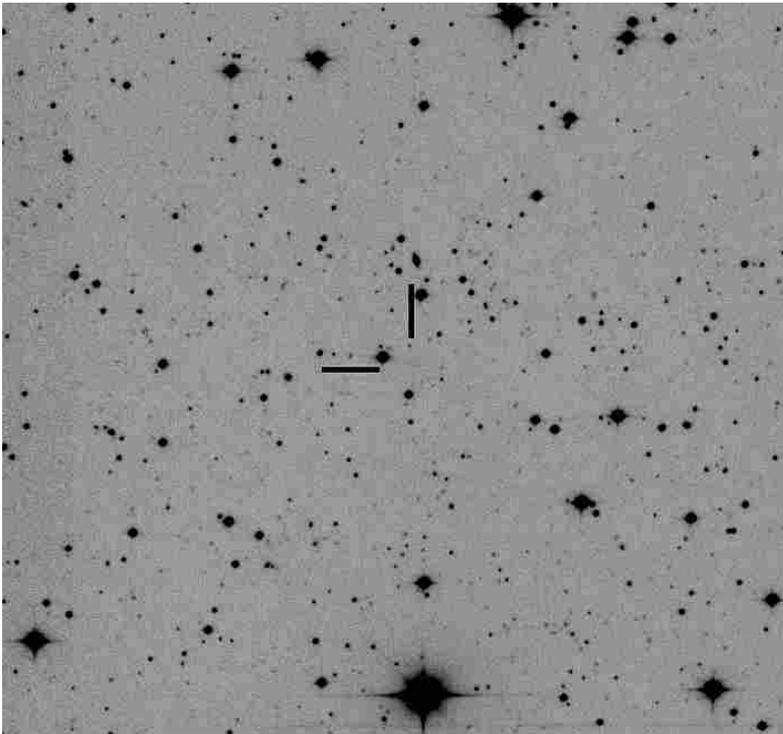

*Fig.3. Supposed new member of the LeoI group, NGC3368dwTBG. The image fragment is spanning 22 arcmin across. North is on top right and East is on top left.*

*3.4. NGC 3521 = KIG 461.* This is an isolated Sbc galaxy at a distance of 10.7 Mpc[8]. Its deep images obtained by W. Probst and R. Polzl demonstrate an extremely disturbed periphery of NGC 3521 with a semi-disrupted diffuse companion on the northern side (Figure 4). A new probable satellite of the massive galaxy locates at a separation of 23' or 72 kpc towards South-East

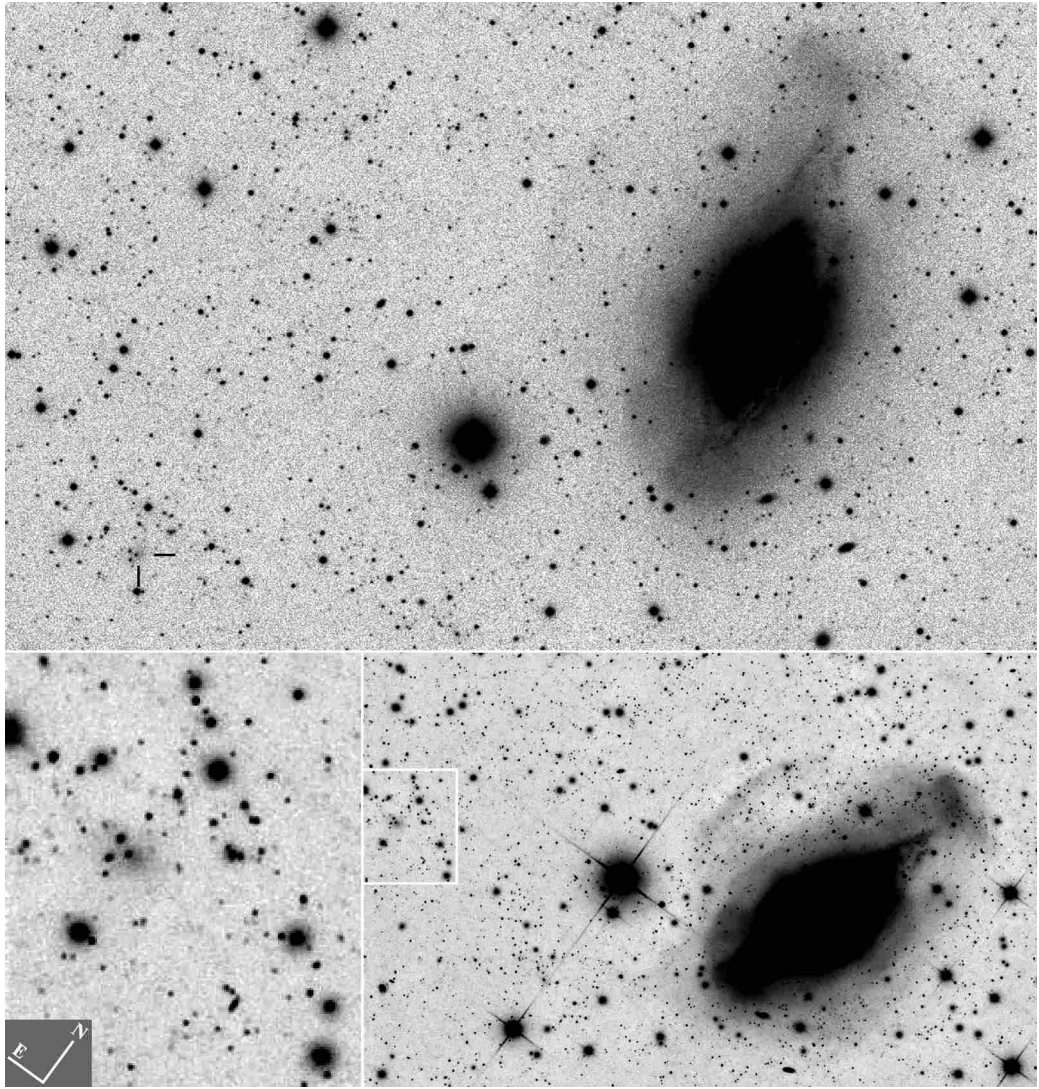

*Fig.4. Mosaic of images of the NGC 3521 having a very disturbed periphery and a new dwarf with low surface brightness, NGC3521dwTBG.*

*3.5. NGC 4236 = KIG 523.* This is a nearby (D = 4.4 Mpc) Sdm-type orphan galaxy without any known satellites. In the NGC 4236 environs imaged by P.Beisser and G.Kerschuber, one can see only two bluish dwarfs suitable to be the spiral galaxy companions: KK125 near a bright red star and NGC 4236dw1 (Figure 5). We have got a spectrum of the latter object using the 6-meter BTA telescope. Judging by the measured radial velocity $V_h$ = 1463±30 km/s, this dwarf galaxy is associated not with NGC 4236 but with a scattered background group: NGC 3879, UGC 6764, UGCA 280, and MCG+11-15-20 having radial velocities in the range of (1310 - 1465) km/s. The radial velocity of the brighter dwarf galaxy KK125 remains still unknown.

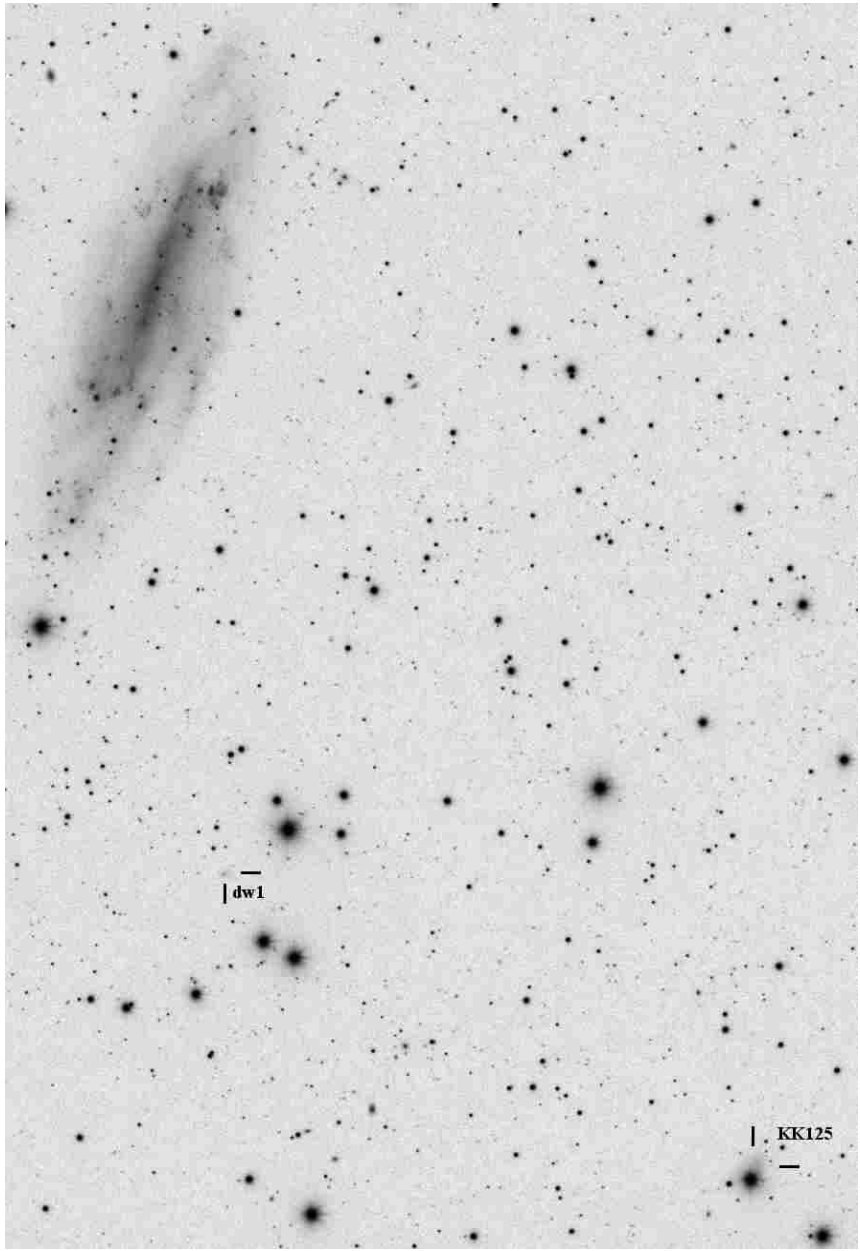

*Fig.5.Spiral galaxy NGC 4236 and two neighbouring dwarfs: KK125 and NGC4236dw1.*

*3.6. NGC 4258 = M106.* Many authors [13, 24-26] investigated the vicinity of this massive spiral galaxy situated at a distance of 7.7 Mpc [8]. Within a projected radius of ~200 kpc around it, there are about 20 candidates to low-surface-brightness satellites. Cohen et al. [27] recently observed four of them with the Hubble Space Telescope and found that only one object, KK132, is the true satellite of NGC 4256, while others belong to the distant background. The reason for this unexpected result can be the fact that NGC 4256 locates at the equator of the Local Supercluster, and its neighbourhood is contaminated with members of other projected groups. In particular, the galaxy groups around NGC 4346 and NGC 4157 have similar radial velocities but distances of about 17 Mpc [28]. A deep image of NGC 4258 surroundings derived by F. Neyer (Figure 6) reveals a new object of very low surface brightness, M106edgeN4217, in contact with the more distant edge-on galaxy NGC 4217, but is likely a satellite of NGC 4258.

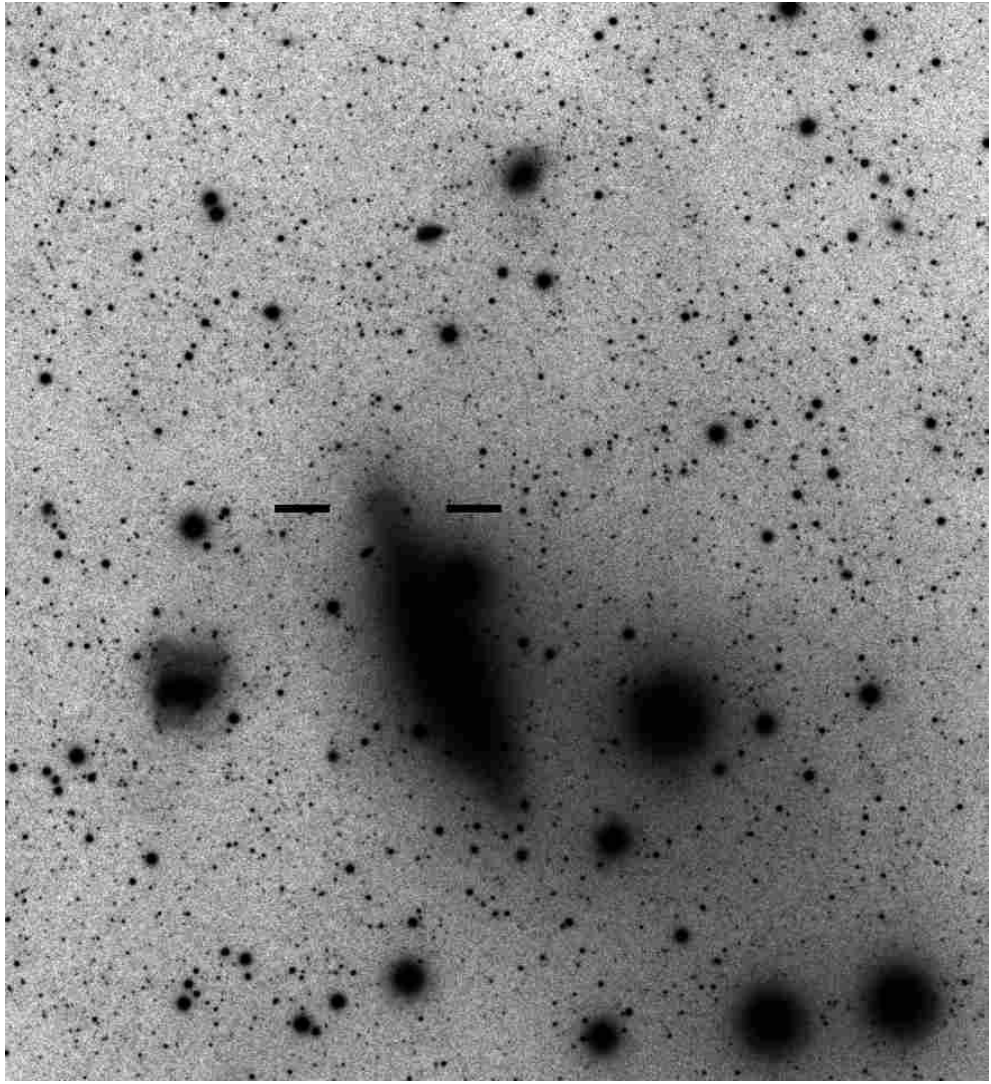

*Fig.6. Eastern vicinity of the spiral galaxy NGC 4258 = M106. The supposed new satellite M106edgeN4217 locates in contact with the distant galaxy NGC 4217 at its northeastern edge. The image fragment size is 27'x 29'.*

3.7. NGC 4594 = M 104 = Sombrero. This remarkable Sa-type galaxy with a prominent bulge, situated at a distance of 9.55 Mpc [29], has the highest luminosity among the Local Volume galaxies. In its vicinity, there are five supposed dwarf satellites of low surface brightness [30]. Only one of them, KKSG30, has been confirmed as the true Sombrero satellite due to its measured radial velocity. Later [17], two more LSB objects were found near Sombrero. A deep image of M104 obtained by S. Kuppers in 2015 (Figure 7) reveals two new dSph galaxies: Sombrero-dwA and Sombrero-dwB, whose morphology allows them to be attributed to true satellites of M104. The first of them has been re-discovered in [17] as NGC4594-DGSAT-2.

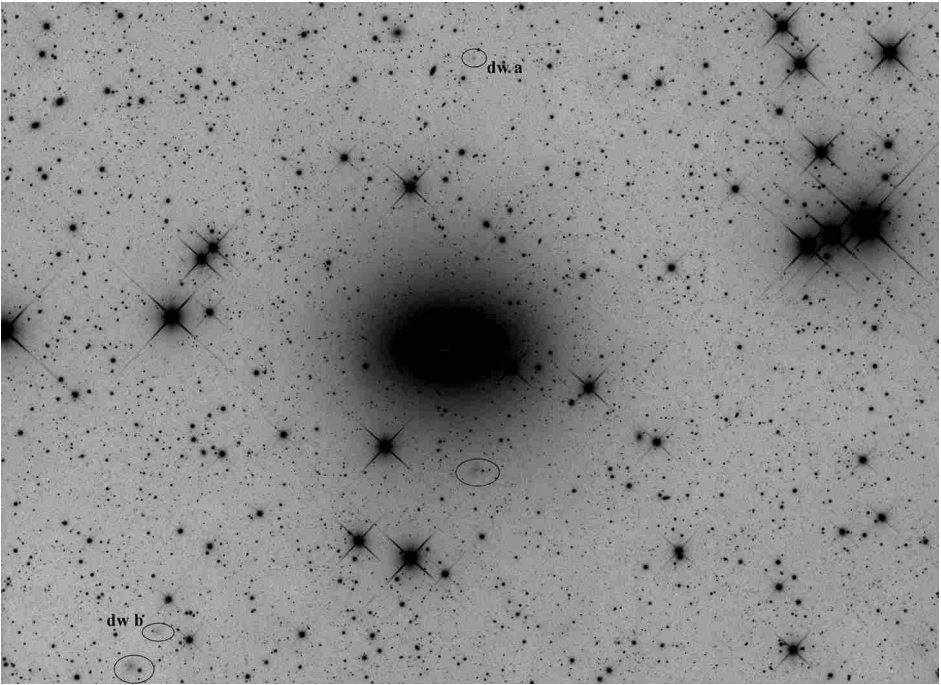

Fig.7. Giant galaxy NGC4594 = M104 is in the image center. Four its diffuse dwarf companions are outlined by ellipses: KKSG32 (below the center), Sombrero-dwTBGa (top), KKSG34 and Sombrero-dwTBGb (bottom left). North is top right and East is to top left.

3.8. NGC 5055 = M63. In a faint periphery of this Sbc galaxy, there is an extended system of low surface brightness stellar streams [31]. The galaxy situated at a distance of 9.0 Mpc [32] has six relatively bright companions. The image of M63 derived by O. Schneider (Figure 8) finds two low-surface-brightness objects: NGC5055dwTBG1 and NGC5055dwTBG2 which can be assigned to the M63 satellites.

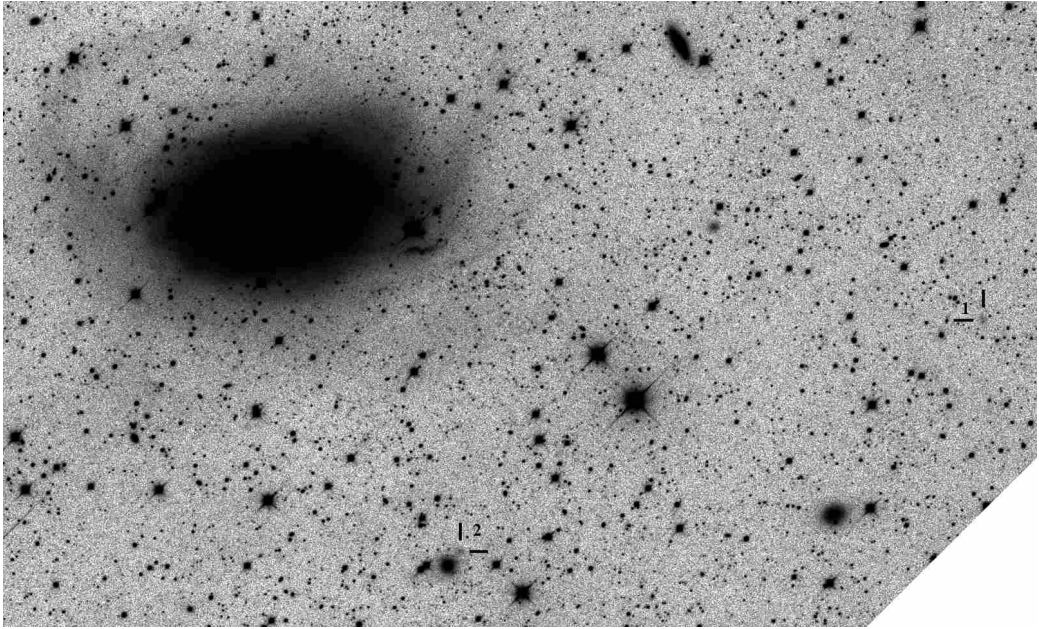

Fig.8. Spiral galaxy NGC5055 = M63 with peripheric stellar streams and two new dwarfs of very low surface brightness: NGG5055dwTBG1 and NGG5055dwTBG2.

*3.9. NGC 6744.* Near this southern Sb galaxy situated at a distance of 9.5 Mpc [32], there are four bright satellites: NGC 6684, IC 4710, IC 4870, ESO 104-044, and 3 supposed companions with low surface brightness: [KK2001]70, [KK2001]71, and [KK2001]72. In the images of NGC 6744 obtained by B. Gludan and S. Kuppers, there are two more objects of very low surface brightness: NGC6744dwTBGa (Figure 9a) and NGC6744dwTBGb (Figure 9b) which can also be the massive galaxy satellites.

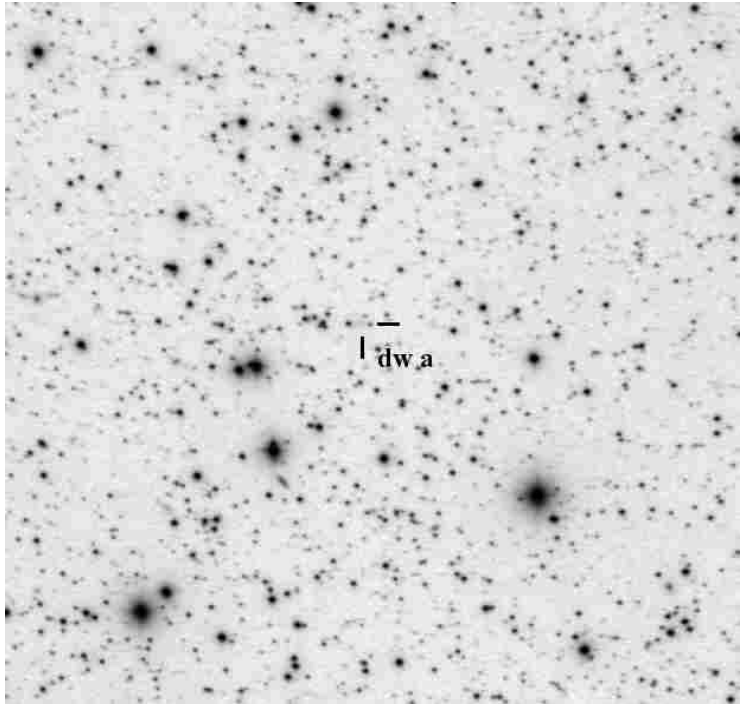

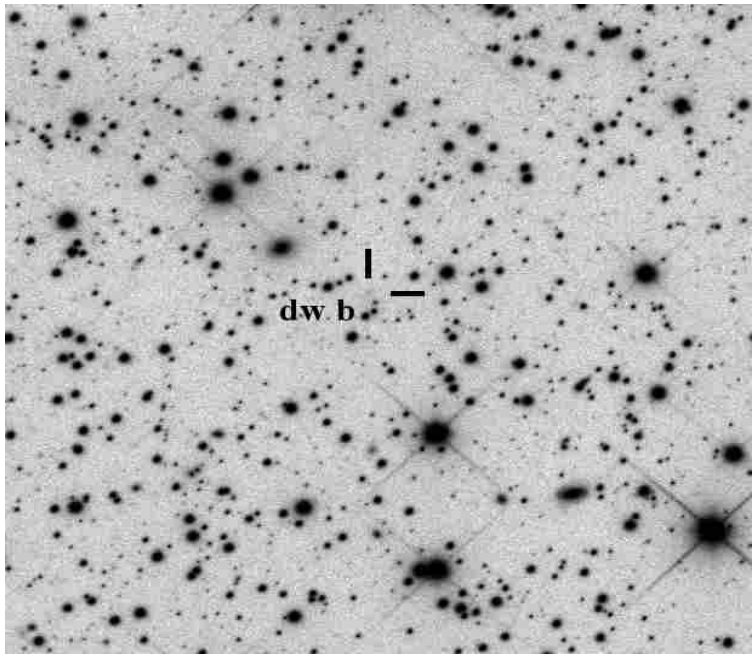

*Fig.9. Supposed new companions of the spiral galaxy NGC6744: dwTBGa (a) and dwTBGb (b). The image fragment sizes are 21'x21' and 11'x11', respectively.*

Table 1. Probable new LSB dwarf galaxies in the Local volume.

| Object | RA(2000.0)Dec | a' | b/a | mFUV mag | B mag | Type | D Mpc | r'p | Rp kpc | A kpc | MB mag | SB |
|---|---|---|---|---|---|---|---|---|---|---|---|---|
| 1 | 2 | 3 | 4 | 5 | 6 | 7 | 8 | 9 | 10 | 11 | 12 | 13 |
| NGC628dwTBG | 013306.8+161505 | 0.92 | 0.78 | >23.0 | 19.0 | Tr-VL | 10.2 | 59 | 174 | 2.7 | -11.0 | 27.5 |
| NGC2337dwTBG1 | 070829.7+443726 | 0.40 | 0.82 | 21.5 | 18.5 | Im-L | 11.9 | 21 | 73 | 1.4 | -11.9 | 25.1 |
| NGC3368dwTBG | 104708.0+111702 | 0.35 | 0.95 | >23.0 | 20.5 | Sph-XL | 10.4 | 33 | 99 | 1.1 | -9.6 | 26.8 |
| NGC3521dwTBG | 110713.1-001115 | 0.34 | 0.92 | >23.0 | 19.5 | Sph-VL | 10.7 | 23 | 72 | 1.1 | -10.6 | 25.8 |
| KK125=HS148 | 121241.4+685533 | 0.80 | 0.58 | 22.0 | 17.4 | Ir-L | 4.4 | 39 | 50 | 1.0 | -10.8 | 25.5 |
| NGC4236dw1 | 121614.3+690625 | 0.38 | 0.77 | 19.9 | 18.7 | Ir-N | 22.2 | 21 | - | 2.5 | -13.0 | 25.2 |
| M106edgeN4217 | 121612.5+470804 | 0.88 | 0.95 | >23.0 | 18.4 | Sph-VL | 7.7 | 30 | 67 | 2.0 | -11.0 | 26.8 |
| Sombrero-dwA | 123951.5-112029 | 0.22 | 0.95 | >23.0 | 20.0 | Sph-L | 9.6 | 17 | 47 | 0.6 | -9.9 | 25.4 |
| Sombrero-dwB | 124112.0-115333 | 0.44 | 0.90 | >23.0 | 19.5 | Sph-L | 9.6 | 24 | 67 | 1.2 | -10.4 | 26.3 |
| NGC5055dwTBG1 | 131218.9+415837 | 0.52 | 0.78 | >23.0 | 19.5 | Tr-VL | 9.0 | 39 | 103 | 1.4 | -10.3 | 26.7 |
| NGC5055dwTBG2 | 131451.1+414330 | 0.48 | 0.90 | >23.0 | 19.5 | Tr-VL | 9.0 | 21 | 56 | 1.3 | -10.3 | 26.5 |
| NGC6744dwTBGa | 190556.0-631619 | 0.42 | 0.95 | >23.0 | 19.0 | Sph-VL | 9.5 | 43 | 120 | 1.2 | -10.9 | 25.8 |
| NGC6744dwTBGb | 191246.8-633949 | 0.22 | 0.85 | >23.0 | 21.0 | Tr-VL | 9.5 | 23 | 64 | 0.6 | -8.9 | 26.3 |

*4. Discussion.* Basic parameters of new dwarf galaxies found by us are presented in Table 1. Its columns contain: (1) object name, (2) equatorial coordinates at J2000.0 epoch, (3, 4) effective angular diameter in arcminutes and axial ratio, (5) apparent FUV magnitude from the GALEX survey [33], (6) apparent B-band magnitude estimated by eye with an accuracy of 0.5 mag via comparison with other objects having photometric data, (7) morphological type of the dwarf galaxy according to classification [8], (8) distance from the Milky Way assuming physical connection between the dwarf and its principal galaxy, (9, 10) angular (arcmin) and linear (kpc) projected separation of the satellite from the main galaxy, (11-13) linear diameter (kpc), absolute B magnitude, and mean surface brightness (mag/sq.arcsec) of the dwarf.

As it follows from these data, the linear diameters of new dwarf galaxies (0.6-2.7) kpc, absolute magnitudes (-8.9- -13.0), and their mean surface brightnesses (25.1 - 27.5) mag/sq.arcsec are typical of the known dwarf spheroidal and dwarf irregular galaxies in the Local Group and other neighbouring groups. The mean linear projected separation of the dwarfs from their main galaxies is 83 kpc that is 2-3 times smaller than the typical halo radius of a Milky Way-type galaxy.

The average values of these parameters: <A> = 1.4 kpc, <MB> = −10.7 mag, <SB> = 26.1 mag/sq.arcsec, and <Rp> = 83 kpc turn out to be very close to the mean parameters for 27 dwarfs found by us before around other bright galaxies in the Local Volume [16]: <A> = 1.3 kpc, <MB> = −10.4 mag, <SB> = 26.1 mag/sq.arcsec, and <Rp> = 73 kpc.

The continuation of taking images of nearby luminous galaxies at small telescopes with the extension of surveying area around them up to 200-300 kpc may lead to the discovery of new dwarf satellites of very low surface brightness.

This work was supported by RFBR grant 18-02-00005.